\documentclass[prb,twocolumn,showpacs,superscriptaddress, amsmath,amssymb]{revtex4}

\usepackage{graphicx}% Include figure files
\usepackage{dcolumn}% Align table columns on decimal point
\usepackage{bm}% bold math
\usepackage[dvips]{color}
\usepackage{color}

\newcommand{\vect}[1]{\mathbf{#1}}

\begin{document}
\preprint{APS/123-QED}

\title{Magnetic phases of quasi-two-dimensional antiferromagnet
on triangular lattice  CuCrO$_2$}

\author{Yu. A. Sakhratov}
\affiliation{National High Magnetic Field Laboratory, Tallahassee, Florida
32310, USA} \affiliation{Kazan State Power Engineering University, 420066
Kazan, Russia}

\author{L. E. Svistov}
\email{svistov@kapitza.ras.ru}
 \affiliation{P. L. Kapitza Institute for
Physical Problems RAS, 119334 Moscow, Russia}

\author{P. L. Kuhns}
\affiliation{National High Magnetic Field Laboratory, Tallahassee, Florida
32310, USA}

\author{H. D. Zhou}
\affiliation{National High Magnetic Field Laboratory, Tallahassee, Florida
32310, USA} \affiliation{Department of Physics and Astronomy, University of
Tennessee, Knoxville, Tennessee 37996, USA}

\author{A. P. Reyes}
\affiliation{National High Magnetic Field Laboratory, Tallahassee, Florida
32310, USA}

\date{\today}

\begin{abstract}
We have carried out $^{63,65}$Cu NMR spectra measurements in magnetic field up
to about 45~T on single crystal of a multiferroic triangular
antiferromagnet CuCrO$_2$. The measurements were performed for magnetic fields
aligned along the crystal $c$-axis. Field and temperature evolution of the spectral shape
demonstrates a number of phase transitions. It was found that the 3D
magnetic ordering takes place in the low field range ($H\lesssim15$~T). At higher fields
magnetic structures form within individual triangular planes whereas the spin directions of the magnetic ions from neighboring planes are not correlated. It is established that the 2D-3D transition is hysteretic in field and temperature. Lineshape analysis reveals several possible magnetic structures existing within individual planes for different phases of CuCrO$_2$. Within certain regions on the magnetic H-T phase diagram of CuCrO$_2$ a 3D magnetic ordering with tensor order parameter is expected.

\end{abstract}


\maketitle

\section{Introduction}

The problem of an antiferromagnet on a triangular planar lattice has been
intensively studied
theoretically.~\cite{Kawamura_1985,Korshunov_1986,Anderson_1987,Plumer_1990,Chubukov_1991}
The ground state in the Heisenberg and XY models is a ``triangular'' planar
spin structure with three magnetic sublattices arranged 120$^\circ$ apart. The
orientation of the spin plane is not fixed in the exchange approximation in the
Heisenberg model. For the simplest semiclassical model of triangular structure with single antiferromagnetic exchange interaction between nearest neighbors the applied static field does not remove the degeneracy of
the spin configurations. Therefore the usual small corrections such
as quantum and thermal fluctuations, and relativistic interactions in the
geometrically frustrated magnets play an important role in the formation of the
equilibrium state.~\cite{Korshunov_1986,Chubukov_1991,Rastelli_1996} Interests in
triangular antiferromagnets are fueled by a rich variety of exotic
phases which can be realized in such systems. Such model systems can be tested
experimentally in three-dimensional (3D) crystals, where magnetic ions are located in the
triangular lattice sites of crystallographic planes. If the in-plane interactions
strongly exceed inter-plane interactions we can expect that such quasi-two-dimensional (quasi-2D)
magnet will demonstrate the features typical for 2D models. Various
realizations of quasi-2D antiferromagnets on triangular lattice are discussed
in reviews (Refs. \onlinecite{Collins_1997, Starykh_2015}).

CuCrO$_2$ is an example of quasi-2D antiferromagnet ($S=3/2$) with
triangular lattice structure. Early neutron scattering experiments reveal that the electronic spin structure in CuCrO$_2$ is in a planar 120-degree configuration below $T_c\approx 24$~K,~\cite{Kadowaki_1990} with a disorder caused by the frustration of the inter-plane exchange bonds.
More recently, neutron scattering investigations~\cite{Poienar_2009} in CuCrO$_2$
single crystals detected a 3D planar magnetic order with
incommensurate wave vector that slightly differs from the wave vector of a
commensurate 120-degrees structure. The magnetic ordering is accompanied by a
simultaneous crystallographic distortion~\cite{Kimura_JPSJ_2009} of the regular
triangular lattice and by the appearance of an electrical polarization.
According to neutron scattering experiments~\cite{Poienar_2010} the interaction between nearest Cr$^{3+}$ ions within triangular plane is strongest. This in-plane interaction is more than 20 times larger than the frustrated inter-plane interaction. Thus, CuCrO$_2$ can be considered as an example of quasi-2D antiferromagnet.

We present a NMR study of the low temperature magnetic structure
of CuCrO$_2$ in fields up to 45~T aligned along the crystallographic $c$-axis.
Such field direction corresponds to orientation within the spin plane of the spiral structure in low field range.
The highest field in our experiments is about 1/6 of the
saturation field which can be estimated using susceptibility value as $\mu_0H_{sat}\approx 280$~T. Analysis of temperature and field
evolution of NMR spectral shape reveals the magnetic H-T phase diagram
of CuCrO$_2$ and suggests the realized magnetic phases. In most respects, the main features of
the observed phase diagram are consistent with the phase diagram obtained from electric
polarization measurements measured in pulsed magnetic fields.\cite{Zapf_2014, Lin_2014} The
observed spectra at low magnetic field ($\mu_0 H\lesssim15$~T) can be well described
by a 3D magnetic structure as detected by neutron experiments at zero field,~\cite{Poienar_2010} whereas
the NMR spectra observed at higher fields indicate the loss of the 3D magnetic
ordering of CuCrO$_2$. In this case the magnetic state of CuCrO$_2$ can be modeled
as a number of randomly stacked independently ordered 2D triangular magnetic layers.
The transition from 3D to 2D state exhibits hysteretic behavior in both field and temperature. Surprisingly, the magnetic phase identified with NMR as 2D state at the same H-T region also demonstrates magnetically driven electric polarization.\cite{Zapf_2014}
These facts indicate that the high field magnetic structure is, in fact, a 3D-polar phase with tensor order parameter, akin to Andreev and Grischuk's $p$-type nematic phase.\cite{Andreev_1984}

\section{Crystal and magnetic structure}

\begin{figure}[b!]
\includegraphics[width=.7\columnwidth,angle=0,clip]{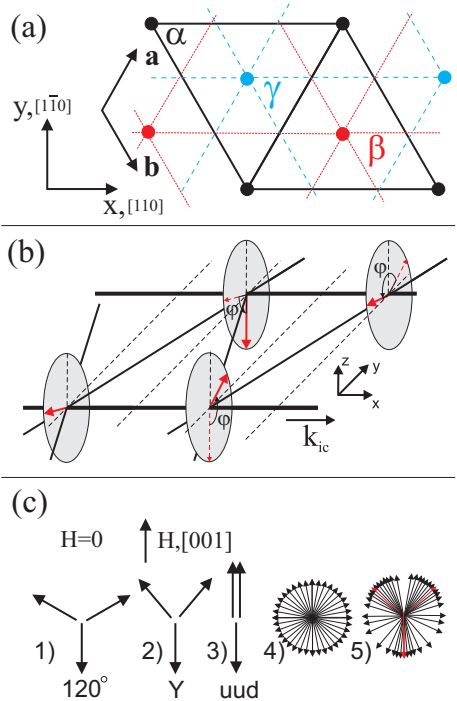}
\caption{(color online) (a) Crystal structure of CuCrO$_2$ projected on the
$ab$-plane. The three layers, $\alpha\beta\gamma$, are the positions of
Cr$^{3+}$ ions. (b) Schema of the spin structure within one triangular plane at zero field (solid red  arrows).
The gray circles show the orientation of the spin plane (110). $\varphi=118.5^\circ$ is a pitch angle of the structure, see text. The incommensurate wavevector $\vect{k}_{ic}$ is collinear with the base of the triangle (thick line). (c) Schema of spin configurations for regular and weakly distorted 2D-triangle structures at low fields.
}
\label{fig:structure}
\end{figure}

The structure CuCrO$_2$ consists of magnetic Cr$^{3+}$ (3$d^3$, $S=3/2$),
nonmagnetic Cu$^+$, and O$^{2-}$ triangular lattice planes,
which are stacked along $c$-axis in the sequence Cr-O-Cu-O-Cr
(space group $R\bar{3}m$, $a=2.98$~\AA{}, $c=17.11$~\AA{} at room
temperature~\cite{Poienar_2009}).
The layer stacking sequences are $\alpha\gamma\beta$, $\beta\alpha\gamma$,
and $\beta\beta\alpha\alpha\gamma\gamma$ for Cr, Cu and O ions, respectively.
The crystal structure of CuCrO$_2$ projected on the $ab$-plane is shown in Fig.\ref{fig:structure}a. The distances between the nearest planes denoted by different Greek
letters for copper and chromium ions and the pairs of planes for oxygen
ions are $c/3$, whereas the distance between the nearest oxygen planes
denoted by the same letters is $(1/3-0.22)c$ ~(Ref.~[\onlinecite{Poienar_2009}]).
No structural phase transition has been reported at temperatures higher than N\'{e}el
ordering temperature ($T > T_c\approx 24$~K). In the magnetically ordered state
the triangular lattice is distorted, so that one side of the triangle becomes
slightly smaller than the two other sides:
$\Delta a / a \simeq 10^{-4}$~(Ref.~[\onlinecite{Kimura_JPSJ_2009}]).

The magnetic structure of CuCrO$_2$ has been intensively investigated by
neutron diffraction experiments.~\cite{Poienar_2009, Kadowaki_1990, Soda_2009, Soda_2010,
Frontzek_2012} It was found that the magnetic ordering in CuCrO$_2$ occurs
in two stages.~\cite{Frontzek_2012, Aktas_2013} At the higher transition
temperature $T_{c1}=24.2$~K, a transition to a 2D ordered state occurs,
whereas below $T_{c2}=23.6$~K, a 3D magnetic order with incommensurate propagation vector
$\vect{k}_{ic}= (0.329,0.329,0)$ along the distorted side of triangular lattice planes~\cite{Kimura_JPSJ_2009}
is established. The magnetic moments of Cr$^{3+}$ ions can be described by the expression
\begin{eqnarray}
\vect{M}(\vect{r}_{i,j})=M_1\vect{e}_1\cos(\vect{k}_{ic}\vect{r}_{i,j}+\Theta)+M_2\vect{e}_2\sin(\vect{k}_{ic}\vect{r}_{i,j}+\Theta),
\label{eqn:spiral}
\end{eqnarray}
where $\vect{e}_1$ and $\vect{e}_2$ are two perpendicular unit vectors
determining the spin plane orientation with the normal vector
$\vect{n}=\vect{e}_1 \times \vect{e}_2$, $\vect{r}_{i,j}$ is the vector to the
$i,j$-th magnetic ion and $\Theta$ is an arbitrary phase. The spin plane
orientation and the propagation vector of the magnetic structure are
schematically shown at the bottom of Fig.\ref{fig:structure}. For zero magnetic field
$\vect{e}_1$ is parallel to $[001]$ with $M_1 = 2.8(2)~\mu_B$, while
$\vect{e}_2$ is parallel to $[1\bar{1}0]$ with $M_2 = 2.2(2)~\mu_B$
(Ref.~[\onlinecite{Frontzek_2012}]). The pitch angle between the neighboring Cr
moments corresponding to the observed value of $\vect{k}_{ic}$ along the
distorted side of triangular lattice planes is equal to 118.5$^\circ$ which is very close to
120$^\circ$ expected for regular triangular lattice planes structure.

Owing to the crystallographic symmetry at $T > T_c$ in the ordered phase ($T < T_c$)
we can expect {\it six} magnetic domains. The propagation vector of each domain can
be directed along one side of the triangle and can be positive or negative. As
reported in Refs.~[\onlinecite{Soda_2010, Svistov_2013, Sakhratov_2014}], the
distribution of the domains is strongly affected by the cooling history of the
sample. For field aligned along [001] axis all six domains are equivalent.

Inelastic neutron scattering data~\cite{Poienar_2010} has shown that CuCrO$_2$
can be considered as a quasi-2D magnet. The spiral magnetic structure is
defined by the strong exchange interaction between the nearest Cr$^{3+}$ ions within
the triangular lattice planes with exchange constant $J_{ab}=2.3$~meV. The inter-planar interactions
are at least one order of magnitude weaker than the in-plane interaction and
frustrated.

Results of the magnetization, electric polarization, ESR and NMR
experiments~\cite{Kimura_PRL_2009, Svistov_2013, Sakhratov_2014} have been
discussed within the framework of the planar spiral spin structure at fields
studied experimentally: $\mu_0H < 14$~T~$\ll \mu_0H_{sat}$.
($\mu_0H_{sat}\approx 280$~T). The orientation of the spin plane is defined by
the biaxial crystal anisotropy. One {\it hard} axis for the normal vector
$\vect{n}$ is parallel to the $c$ direction and the second axis is
perpendicular to the direction of the distorted side of the triangle. The
anisotropy along $c$ direction dominates with anisotropy constant approximately
hundred times larger than that within $ab$-plane resulting from the distortions
of the triangular structure. A magnetic phase transition was observed for the
field applied perpendicular to one side of the triangle ($\vect{H}\parallel
[1\bar{1}0]$) at $\mu_0H_c = 5.3$~T, which was consistently
described~\cite{Kimura_PRL_2009, Soda_2010, Svistov_2013} by the reorientation
of the spin plane from $(110)$ ($\vect{n}\perp\vect{H}$) to $(1\bar{1}0)$
$(\vect{n}\parallel\vect{H})$. This spin reorientation to ``umbrella like'' phase
happens due to the weak susceptibility anisotropy of the spin structure
$\chi_{\parallel}\approx 1.05\chi_{\perp}$, where $\parallel$ and $\perp$
refer to fields parallel and perpendicular to
$\vect{n}$, respectively. Further increase of applied field does not result to any additional
phase transitions up to fields $\approx 60$~T.~\cite{Zapf_2014, Lin_2014}

For fields directed parallel to $c$ axis the magnetic phase diagram is
much more complex. According to electric polarization studies in CuCrO$_2$, at fields
up to 92~T~\cite{Zapf_2014, Lin_2014} the low temperature magnetic structure undergo a number of transitions. The phases realized during the magnetization process are not yet identified and shall be discussed in this paper.

\section{Sample preparation and experimental details}

The sample we used in this experiment is the same as described in ~Ref.~[\onlinecite{Sakhratov_2014}].
Measurements were taken on a superconducting Cryomagnetics 17.5~T magnet, a 30~T resistive magnet and 45~T hybrid magnet at the National High Magnetic Field Laboratory.
All magnets were field sweepable.
For technical reasons zero field cooling of the sample could not be performed
while using 45~T magnet.
$^{63,65}$Cu nuclei (nuclear spins $I=3/2$, gyromagnetic
ratios $^{63}\gamma/2\pi=11.285$~MHz/T, $^{65}\gamma/2\pi=12.089$~MHz/T) were
probed using pulsed NMR technique. The spectra
were obtained by summing fast Fourier transforms (FFT)
or integrating the averaged spin-echo signals as the field was swept through the resonance line.
NMR spin echoes were obtained using $\tau_p~-~\tau_D~-~2\tau_p$ pulse sequences, where the pulse
lengths $\tau_p$ were 1-3~$\mu$s, the times between pulses $\tau_D$ were 15~$\mu$s.
Measurements were carried out in the temperature range $2\leq T \leq 40$~K stabilized with a
precision better than 0.1~K.

\section{Experimental results}

The crystal cell of CuCrO$_2$ contains single copper ion. As a result the
$^{63,65}$Cu NMR spectra for the paramagnetic states consist of two sets of
triplets, corresponding to $^{63}$Cu and $^{65}$Cu isotopes. Each triplet corresponds
to quadrupolar split transitions: a central line ($m_I = +1/2\leftrightarrow -1/2$) and two quadrupole satellite transitions ($m_I = \pm 3/2\leftrightarrow \pm 1/2$).\cite{Sakhratov_2014}
Below the magnetic ordering temperature, the narrow NMR line transforms into a broad spectrum characteristic of incommensurate magnetic structure. The shape of NMR spectrum is defined by the distribution of local magnetic fields on the copper nuclei within the sample. The observed shapes of NMR spectra depend on field, temperature and cooling
history. For different lines in the triplet the spectra measured at the same conditions were found
to be of the same shape. So, we chose the lines which were well distanced from
neighbours to avoid overlapping.

\begin{figure}
\includegraphics[width=0.95\columnwidth,angle=0,clip]{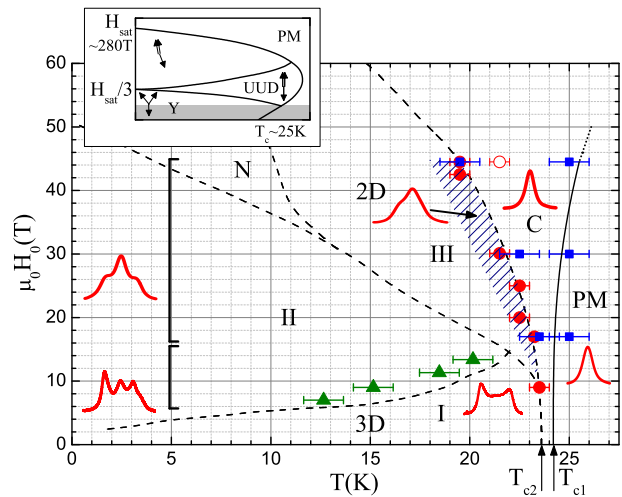}
\caption{(color online)
H-T magnetic phase diagram of CuCrO$_2$. The dashed lines are the boundaries of magnetic phases possessing electric polarization from Ref.\onlinecite{Zapf_2014}. Symbols mark the boundaries between different phases obtained from temperature evolution of $^{63,65}$Cu NMR spectra. Red circles and green triangles represent fields where  transformations of NMR spectra shapes were observed.
Solid red circles denote the points at which the spectral shape is transformed from the paramagnetic phase when the sample is cooled in field. Open red circle corresponds to the spectra taken as the temperature increases. Blue squares show positions of kink like anomalies in temperature dependencies of spin-lattice relaxation rate T$_1^{-1}$(T). Solid line is guide to the eyes. NMR spectra observed in different areas are marked schematically on the phase diagram. Inset is the sketch of phase diagram taken from Ref.~[\onlinecite{Miyashita_1986}]. Shaded areas cover the experimental H-T region within which this work was performed.
}
\label{fig:phasediagram}
\end{figure}

In order to facilitate the discussion connecting the magnetic structure to the NMR spectra, we shall refer extensively to the schematic phase diagram of CuCrO$_2$ shown in Fig.\ref{fig:phasediagram}. The phase transitions observed  in Refs.[\onlinecite{Zapf_2014, Lin_2014}] as anomalies on field dependencies of electric polarization are marked with dashed lines. The two-step transition from paramagnetic state to the ordered phase at zero field are marked with arrows.~\cite{Frontzek_2012, Aktas_2013}

\begin{figure}
\includegraphics[width=0.9\columnwidth,angle=0,clip]{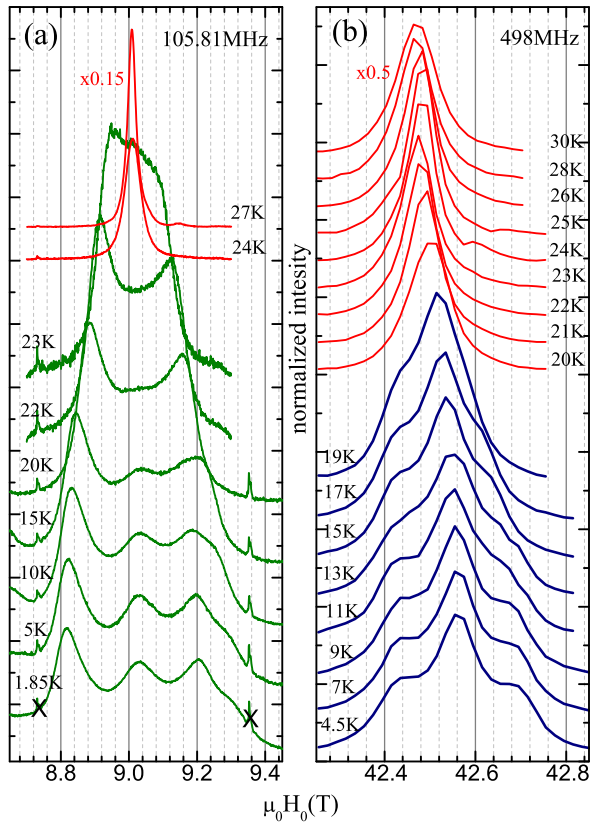}
\caption{(color online)
Typical temperature evolution of $^{63}$Cu NMR spectra ($m_I = +1/2\leftrightarrow -1/2$ transition) for field cooled samples. Data are taken at frequencies a) 105.81~MHz and b) 498~MHz. Color reference to the spectral lineshape as described in the text: red - single peak, green - double horn pattern (both distorted and undistorted), and blue - helmet shape (symmetic and asymmetric).
 The peaks marked with crosses are spurious $^{63,65}$Cu NMR signals from the probe.
}
\label{fig:temperatureevol}
\end{figure}

The temperature evolutions of $^{63}$Cu NMR spectra measured at frequencies 105.81~MHz, 498~MHz cooled in field from the paramagnetic state are shown in the Fig.\ref{fig:temperatureevol}. The spectra shown in Fig.\ref{fig:temperatureevol}a ($\mu_0H\approx 9$~T) is typical for spectra measured at fields below 15~T.
At high temperatures, $T>T_{c2}$, each spectrum is characterized by a single peak, as expected for paramagnetic phase. On the other hand, below $T_{c2}$, the spectra demonstrate a two-horn pattern characteristic of incommensurate structure, with a low-field peak more intense than the high-field peak.
At temperatures below 20~K the two-horn spectrum becomes distorted: an additional maximum at the middle of the NMR spectrum appears and the high field maximum splits into two.

For fields above 15~T (and up to 45~T), the typical temperature evolution of the spectra is shown in Fig.\ref{fig:temperatureevol}b (showing data for $\mu_0H=42.5$~T). The spectra evolves from a single-peak line for $T\gtrsim20$~K to an asymmetric helmet-shaped line which becomes more symmetric as the temperature is lowered. In the main panel of Fig.\ref{fig:phasediagram}, the cross-hatched region in the phase diagram indicates where the asymmetric helmet lineshape was observed.

\begin{figure}
\includegraphics[width=0.9\columnwidth,angle=0,clip]{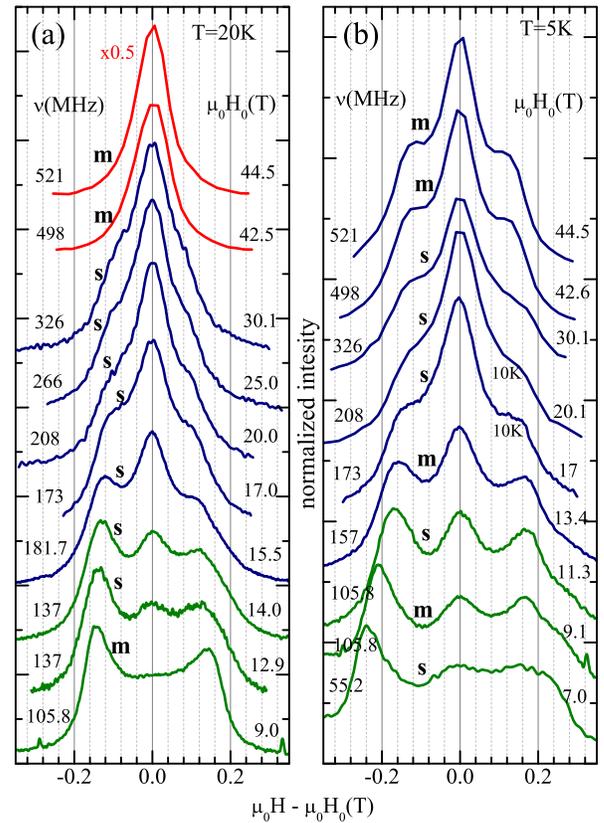}
\caption{(color online)
Field evolution of $^{63}$Cu NMR spectra measured
at temperatures a) 20~K and b) 5~K after field cooling. The spectra are shifted by the values $\mu_0H_0$ indicated to the right of each line. Color identifiers are the same as in Fig.\ref{fig:temperatureevol}. Symbols $m$ and $s$ correspond to the main ($m_I = +1/2\leftrightarrow -1/2$) and high-field satellite ($m_I = +3/2\leftrightarrow +1/2$) transitions, respectively.
}
\label{fig:fieldevol}
\end{figure}

Fig.\ref{fig:fieldevol} shows NMR spectra measured at $T=20$~K and 5~K at different fields. All spectra were measured while the sample is cooled in field (FC). The spectra in Figs.\ref{fig:temperatureevol},\ref{fig:fieldevol}  are differentiated by different colors: spectra with single peaks are red colored, two-horn and distorted two-horn shaped spectra are green colored. NMR spectra with helmet shape or distorted helmet shape spectra are blue colored. The spectrum transformation from single-peaked line-shape to two-horn or helmet-like one is sharp (see Fig.\ref{fig:temperatureevol}). These transitions are marked on the phase diagram by red circles. The error bars at these points are defined by the temperature step between measured spectra.

The transition from the single-peak line to a broad line is hysteretic at high fields. At 44.5~T, the transition temperature upon cooling differs from the transition upon heating by 2~K. The open red circle on the phase diagram at 44.5~T corresponds to the transition observed while warming up. At fields below 17~T, the difference in the transition temperatures between cooling and warming is not more than the temperature steps at which the data were taken.

\begin{figure}
\includegraphics[width=0.9\columnwidth,angle=0,clip]{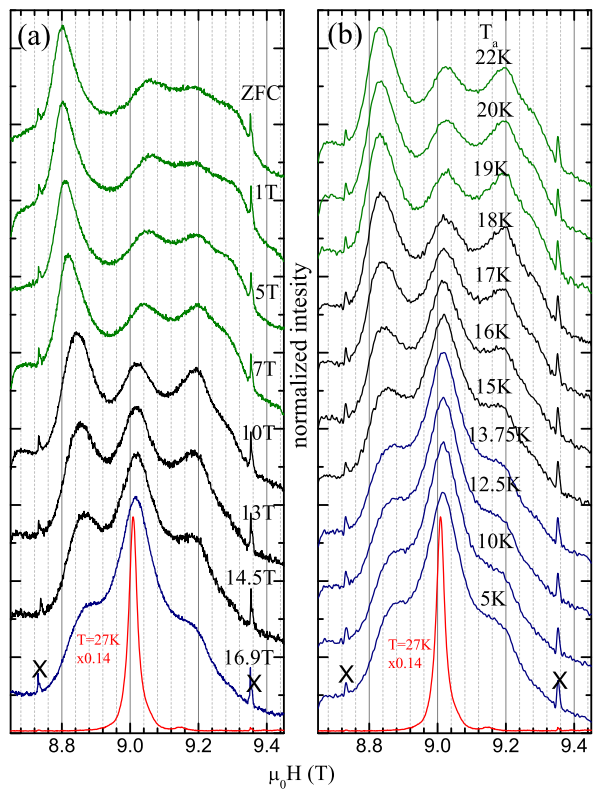}
\caption{(color online)
$^{63}$Cu NMR spectra ($m_I = +1/2\leftrightarrow -1/2$ transition), $\nu=105.81$~MHz, $T=5$~K. The peaks marked with crosses are spurious $^{63,65}$Cu NMR signals from the probe. Color identifiers are the same as in Fig.\ref{fig:temperatureevol}.
The black lines demonstrate the spectra which can be considered as a combination of low field distorted two-horn shaped spectrum and high field helmet-like shaped spectrum.
(a) Series of spectra obtained after cooling in different fields from 40~K to 5~K.
(b) Series of spectra with different annealing temperatures. The first spectrum at the bottom was measured after cooling the sample from 40~K to 5~K in the field $H=16.9$~T. Before other measurements the sample was first annealed at temperature $T_a$ and field $H = 9$~T during 10 minutes then the temperature was set back to 5~K and the spectrum was obtained.
}
\label{fig:FCandTstar}
\end{figure}

The shapes of the NMR spectra at low temperatures depend on the field at which the sample was cooled. Spectra measured at 105.8~MHz and $T=5$~K after field cooling from 40~K to 5~K is shown in the Fig.\ref{fig:FCandTstar}a.

As elaborated in Ref.[\onlinecite{Sakhratov_2014}] the ``double-horn'' and ``helmet'' shaped NMR spectra in CuCrO$_2$ are specific for incommensurate magnetic structures within triangular planes with  inter-plane order and disorder respectively. Green triangles on the phase diagram denotes the boundary between H-T regions where double-horn spectra (i.e. 3D order) is realized independently on cooling procedure and the region where the NMR spectra (or out of plane ordering) depends on  the cooling history. The region with hysteresis is marked with II on the phase diagram. The green points were obtained with the experimental procedure demonstrated in Fig.\ref{fig:FCandTstar}b. First,
the system was prepared by cooling the sample from $T=40$~K to $T=5$~K at a fixed field of 16.9~T and then reducing the field to 9~T. A typical spectrum obtained after such procedure has a helmet shape demonstrated by the lowest blue spectrum in both Fig.\ref{fig:FCandTstar}a (labeled ``16.9~T'') and Fig.\ref{fig:FCandTstar}b (labeled ``5~K'').
While keeping the field at 9~T, the temperature of the sample is then raised to some temperature $T_a$ where it is annealed for 10 minutes before cooling back down to 5~K, at which point the spectrum is again obtained. Fig.\ref{fig:FCandTstar}b shows the evolution of the 5~K spectrum as the annealing temperature is increased. It is evident that at 9~T, the 2D-3D transition, indicated by a transition from helmet to distorted double-horn lineshape, takes place at $T_a=15\pm1$~K. This transition point is represented by a green triangle on the phase diagram.

\begin{figure}
\includegraphics[width=1\columnwidth,angle=0,clip]{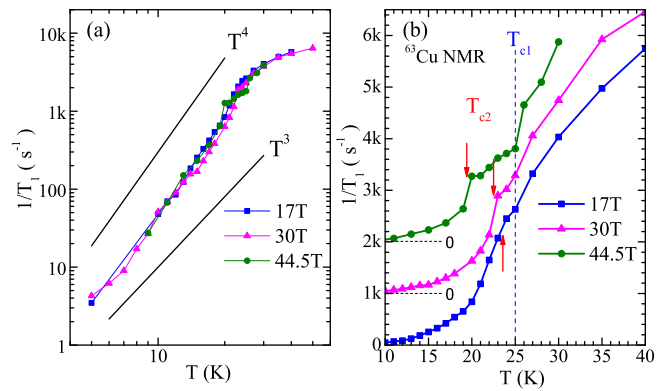}
\caption{(color online)
Temperature dependence of $^{63}$Cu spin-lattice relaxation rate at fields 17, 30 and 44.5~T.
}
\label{fig:t1}
\end{figure}

Not all phases can be differentiated by the shape of NMR spectra. So, within the high temperature phase between $T_{c1}$ and $T_{c2}$ single line shaped spectra were observed, as in the paramagnetic phase. To determine this transition we measured the temperature dependence of spin-lattice relaxation rate at fields 17, 30 and 44.5~T from 40~K to 5~K, shown in Fig.\ref{fig:t1}.
$T_1$ was extracted using multi exponential expression.\cite{Suter_1998}

The temperature dependence of the relaxation rate $T_1^{-1}$ at $T<T_{c2}$ approximately follows a power law behavior with an exponent equal to $3.5\pm0.5$ (Fig.\ref{fig:t1}a).
In Fig.\ref{fig:t1}b, anomalous discontinuities in the behavior of $T_1^{-1}$ (positions marked with arrows) are observed and they are plotted with blue squares in Fig.\ref{fig:phasediagram}.
Note that at temperatures below $T_{c2}$ we did not find any singularity corresponding to the transition to the low temperature high field phase observed in pulse field experiments \cite{Zapf_2014}
identified as phase N in Fig.\ref{fig:phasediagram}.

From these experimental observations and referring to Fig.\ref{fig:phasediagram}, we summarize the phase diagram as follows: 1) There are two regions, identified as C and PM where the NMR spectra are single peaked. 2) Two regions, I and III, are identified with  either two-horn or helmet shaped spectra. These phases are separated by 3) a broad region II, where the spectra demonstrate hysteretic behavior. The spectral shape in this region is a subtle combination of those in I and III and is determined by cooling history. This large hysteresis was also observed in pulsed fields experiments.\cite{Zapf_2014}

\section{Discussion}

The dominant interactions, which govern the magnetic state of CuCrO$_2$, are exchange interactions within triangular $ab$-plane.\cite{Poienar_2010} The intra-plane exchange interaction between spins of nearest chromium ions is at least 20 times larger than the frustrated inter-plane exchange interactions. We shall first discuss the magnetic phases within the model of a 2D antiferromagnet on the regular  triangular lattice. For such a model, in the limit of large spins, the magnetic ground state is expected to be planar three-sublattice 120-degree structure. The spin plane orientation in exchange approximation is arbitrary. This degeneracy may be lifted by taking into account the relativistic interactions with the crystal environment. The orientation of the spin plane in CuCrO$_2$ in the ordered state, i.e. along (110), is defined by two ``easy'' axes of anisotropy: a strong axis [001] perpendicular to triangular planes and another weak axis within the triangular planes $[1\bar{1}0]$. This model can be described by the model Hamiltonian:

\begin{eqnarray}
\textit{H}=\sum_{i,j} \it{J}\vect{S}_i\vect{S}_j+\sum_{i} (\frac{1}{2}A_zS_{i,z}^2+\frac{1}{2}A_yS_{i,y}^2),
\label{eqn:Hamiltonian}
\end{eqnarray}
where the exchange integral between nearest spins: $J\simeq 26.6$~K, \cite{Poienar_2010} and the anisotropy parameters are: $A_z\simeq -0.8$~K, $A_y\simeq -0.0075$~K. \cite{Yamaguchi_2010, Svistov_2013}

The in-plane anisotropy parameter is less than 1~\% of that of the out-of-plane anisotropy. That means, that the model with one ``easy'' axis can be a good approximation. The ground magnetic state in such a model is the so-called three sublattice Y-phase, see Fig.\ref{fig:structure}c1.\cite{Miyashita_1986, Chubukov_1991} This phase possesses spontaneous magnetic moment and at $H=0$ two domains are allowed.
The spontaneous magnetization of three-neighbor magnetic moments is evaluated from the exchange and anisotropy parameters of CuCrO$_2$ as $3\cdot 10^{-3}\mu$, where $\mu$ is magnetic moment of chromium ion.\cite{Miyashita_1986, Chubukov_1991}
The application of magnetic field along anisotropy axis removes the degeneracy. The magnetization of the sample grows monotonically with field up to 1/3 of the saturated magnetization. Such value of magnetization is expected to be maintained in some field range in the vicinity of the value 1/3H$_{sat}$. In this field range the collinear up-up-down (UUD) phase is expected. The field range of UUD phase is stabilized by uniaxial anisotropy and thermal fluctuations. The thermal fluctuations in the vicinity of T$_N$ for anisotropic model can stabilize UUD phase even at zero field.\cite{Miyashita_1986} This fact probably explains the two-stage transition from paramagnetic to ordered phase in CuCrO$_{2}$.

The H-T phase diagram of 2D Heisenberg antiferromagnet on the triangular lattice  from Ref.~[\onlinecite{Miyashita_1986}] is shown as insert to Fig.\ref{fig:phasediagram}. The field range studied in our experiments is shaded.
The field axis was scaled using the value of exchange integral $J= 26.6$~K, which corresponds to $\mu_0H_{sat}\approx 280$~T. Thus within the 2D regular triangular antiferromagnetic model for field range $H<45$~T we can expect Y-phase at low temperature, collinear UUD phase and paramagnetic phase at higher temperatures.
Note, that the magnetic ordering temperature of CuCrO$_2$($\approx 24$~K) is in a good agreement with T$_N=0.51JS(S+1)\approx25$~K, evaluated within 2D XY-model of triangular antiferromagnet.\cite{Landau_1985}

According to Ref.~[\onlinecite{Marchenko_2014}] the main properties of antiferromagnetic CuCrO$_2$ have natural explanation based on Dzyaloshinski-Landau theory of magnetic phase transitions.
Firstly, let us consider the properties of CuCrO$_2$ which can be explained by strongest exchange couplings.
The crystal structure of CuCrO$_2$ allows for the Lifshitz invariant which couples spins of neighboring triangular planes and explains the helicoidal spin structure with incommensurate wave vector. The proximity of the wave vector of magnetic structure for CuCrO$_2$, (0.329,0.329,0),  to the wave vector of a simple 120-degree structure (1/3,1/3,0) demonstrates the smallness of Lifshitz invariant compared with intra-plane exchange interaction. The difference of the pitch angle of the spins of the neighboring magnetic ions versus 120 degrees causes a nonzero magnetic moment on every triangle of the structure rotating within spin plane along the wave vector direction with period approximately equal to hundred sides of triangle structure (see Eq.\ref{eqn:spiral}). The value of noncompensated rotated moment for every three neighbor magnetic ions is calculated to be 0.045$\mu$, where $\mu$ is the magnetic moment of chromium ion. This value is approximately 15 times larger than the magnetic moment expected due to uniaxial anisotropy. The applied magnetic field will distort the triangles so that the total magnetic moments are aligned along energetically preferable field direction. We can expect that the magnetic field can induce the nonharmonic distortion of magnetic structure which stabilizes the preferable Y-like configuration because of its energetic favorability. At high enough fields a transition to commensurate phases is expected.

The inter-plane interaction in CuCrO$_2$ is strongly frustrated and as a result, 3D magnetic ordering with the propagation vector (0.329,0.329,0) observed in the experiment can not be established due to usual exchange interactions between two spins of neighbor triangular planes. 3D ordering can be established, according to Ref.[\onlinecite{Marchenko_2014}], due to the symmetry invariants with at least six spin operators of ions originating from two or three neighbor triangular planes. At the same time, 3D ordering of vector $\boldsymbol{\eta}=\vect{M}(\vect{r}_{i,j})\times \vect{M}(\vect{r}_{i+1,j})$ (see Eq.\ref{eqn:spiral}) can be caused by other invariants with a smaller number of spin operators (namely four). Due to this fact it is possible that the vector $\boldsymbol{\eta}$ will be ordered at a temperature higher than the usual 3D spiral magnetic ordering temperature. The magnetic structure with order parameter $\boldsymbol{\eta}$ can be described by  Eq.\ref{eqn:spiral} with random phases $\Theta$ for different triangular planes.

Symmetry analysis of relativistic interactions in CuCrO$_2$ explains the existence of electric polarization proportional to vector $\boldsymbol{\eta}$ of magnetic structure.\cite{Marchenko_2014} For 3D-ordered spiral phase such polarization was observed experimentally.\cite{Kimura_PRB_2008,Kimura_PRL_2009} It is important that the electric polarization is expected also for magnetic phase with tensor order parameter $\boldsymbol{\eta}$.\cite{Marchenko_2014}

To identify the magnetic phases occuring in CuCrO$_2$, NMR spectra for different model structures were calculated.
In our calculation we assume that the local magnetic field at Cu sites is the sum of the long-ranged dipole field $\vect{H}_{dip}$ and the transferred hyperfine contact field produced by the nearest Cr$^{3+}$ moments. The details of this calculation can be found in our previous study.~\cite{Sakhratov_2014} Below we shall list the model phases that we used to explain the observed shapes of the NMR spectra of CuCrO$_2$ at fields aligned along [001].

\begin{figure}
\includegraphics[width=0.9\columnwidth,angle=0,clip]{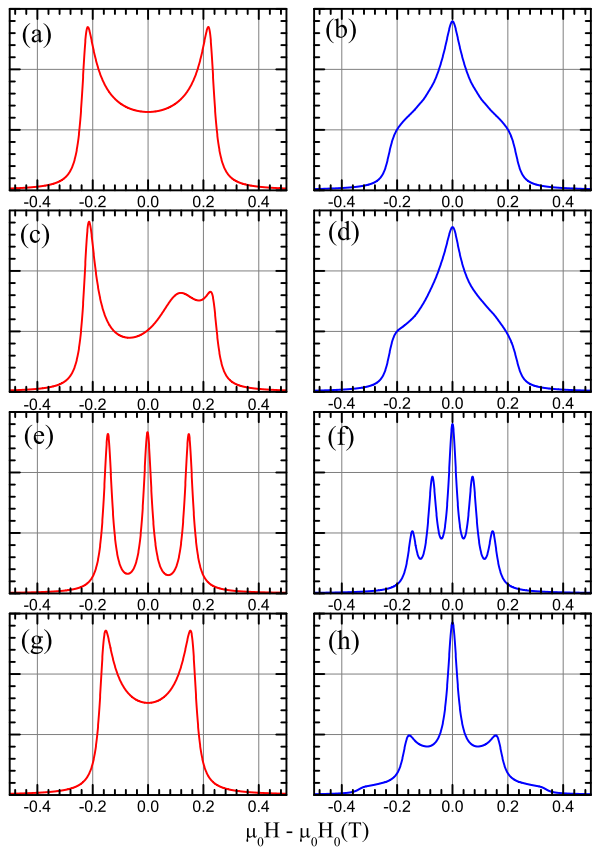}
\caption{(color online)
Simulated NMR spectra corresponding to different models discussed in the text, $M_1=M_2=3\mu_B$, individual linewidth $\delta=18$~mT.
(a),(b),(c),(d),(e),(f) spiral plane (110); (g),(h) spiral plane (001);
(a) 3D incommensurate structure; (b) 3D-polar and 2D incommensurate structures;
(c) 3D incommensurate structure with distortion ($C=0.04$);
(d) 3D-polar and 2D incommensurate structures with distortion ($C=0.01$);
(e) 3D commensurate Y structure; (f) 2D commensurate Y structure;
(g) 3D and 3D-polar incommensurate spin-flopped structures;
(h) 2D incommensurate spin-flopped structure.
}
\label{fig:simul}
\end{figure}

1. Paramagnetic and collinear UUD structures.
\newline
Effective fields generated by magnetic environment on each copper ion are identical.
As a result, one single peaked NMR spectrum is expected. The position of the line is defined by the total magnetization.

2. 3D commensurate Y structure.
\newline
This is described by Eq.\ref{eqn:spiral} with propagation vector (1/3,1/3,0). The initial phase $\Theta$ measured from $\vect{e}_1\parallel\vect{z}$ can be equal to $60^\circ$, $180^\circ$ or $300^\circ$. The choice of $\Theta$ determines one of three possible magnetic domains. The computed spectrum for such 120-degree Y structure (see Fig.\ref{fig:structure}c1) is given in Fig.\ref{fig:simul}e. The commensurate 120-degree Y phase can be deformed due to applied field: the moments of two of three sublattices are tilted towards field direction (Fig.\ref{fig:structure}c2). In this case simulated spectrum consists of three lines of equal intensity as for 120-degree Y structure.

3. 2D commensurate Y structure.
\newline
This is described by Eq.\ref{eqn:spiral} with 2D propagation vector (1/3,1/3).
The magnetic domain which is established within every $ab$-plane was chosen arbitrary. This was achieved by random selection of $\Theta$ among the values $60^\circ$, $180^\circ$, $300^\circ$. The computed spectrum consists of 5 lines, see Fig.\ref{fig:simul}f.

4. 3D incommensurate structure.
\newline
This is described by Eq.\ref{eqn:spiral} with propagation vector (0.329,0.329,0), see Fig.\ref{fig:structure}c4.
The initial phase $\Theta$ can be any, but the same for all triangular planes. The result is shown in
Fig.\ref{fig:simul}a. Such shape with two maxima at the boundaries is typical for incommensurate structures.

5. 3D incommensurate distorted structure.
\newline
For this structure we introduced in Eq.\ref{eqn:spiral} a small nonlinearity in phase:
\begin{eqnarray}
\Phi=(\vect{k}_{ic}\vect{r}_{i,j}+\Theta) + C\cdot \sin{(3(\vect{k}_{ic}\vect{r}_{i,j}+\Theta))},
\label{eqn:newphase}
\end{eqnarray}
where $0<C<0.35$ defines the amplitude of unharmonicity.
Such unharmonicity  tilts the spins within spiral towards to the Y phase (phase angles $60^\circ$, $180^\circ$, $300^\circ$ measured from $\vect{e}_1\parallel\vect{z}$) in contrast to upside down Y phase (phase angles $0^\circ$, $120^\circ$, $240^\circ$), see Fig.\ref{fig:structure}c5. The result is shown in Fig.\ref{fig:simul}c.

6. 3D-polar and 2D incommensurate structures.
\newline
The incommensurate phases, distorted and undistorted, were simulated for two types of disorder. In the first case parameters $\Theta$ in all triangular planes of the structure were chosen arbitrarily (see Eqs.\ref{eqn:spiral},\ref{eqn:newphase}), whereas the direction of the rotation of spins within the triangular planes and, therefore, the sign of vector $\boldsymbol{\eta}$, was set to be the same. We shall assign this phase as 3D-polar ordered phase with order parameter $\boldsymbol{\eta}$.
For the second type of disorder both $\Theta$ and sign of $\boldsymbol{\eta}$ for all triangular planes were arbitrary. We shall assign this phase as 2D-ordered phase. The 2D long range order is possible only at zero temperature. Nevertheless, because the NMR experiment is sensitive only to the nearest magnetic neighborhood of the nuclei, we can expect that the spectra obtained within this model will also describe the magnetic structure with short ranged spiral correlations with different signs of $\boldsymbol{\eta}$, which are expected for 2D triangular structure.\cite{Korshunov_2006} The shapes of simulated spectra for both types of disorder were identical.
The result for undistorted 3D-polar and 2D phases is in Fig.\ref{fig:simul}b, for distorted ones is in Fig.\ref{fig:simul}d.
The characteristic feature of spectra from the magnetic structure with partial disorder is the presence of strong central maximum on the spectra.

7. Spin-flopped umbrella-like structures.
\newline
This is described by Eq.\ref{eqn:spiral} with $\vect{n}\parallel\vect{z}$ for 3D,  3D-polar and 2D phases.
3D and 3D-polar structures both resulted in an identical double horn pattern (Fig.\ref{fig:simul}g).
2D structure resulted in a spectrum with 5 characteristic maxima (Fig.\ref{fig:simul}h).

From comparison of experimental data with the models' simulations we can exclude from consideration commensurate and spin-flopped structures (Fig.\ref{fig:simul}e,f,g,h).
These phases were suggested from microscopic model studied numerically in Ref.~[\onlinecite{Lin_2014}].
The spectra with 3 and 5 maxima were not observed in experiment. Double horn shaped spectra were observed only at low fields, where the umbrella-like structure with $\vect{n}\parallel\vect{c}$ is certainly not realized.

As a result of modeling we suggest the following magnetic structures realized within the studied H-T region.
In all fields below $H\approx45$~T at temperatures below the red solid symbols on the phase diagram an incommensurate spiral phase is established within individual planes. In region I the system is 3D ordered. At higher fields (region III) the 3D-polar or 2D structure is established. Within the broad region II on the phase diagram the spectra exhibits hysteresis where the inter-plane ordering essentially depends on cooling history. At high temperatures $T\gtrsim20$~K the transition from I to III has no field hysteresis. The field hysteresis grows drastically at lower temperatures.

The area on phase diagram bounded by two lines between $T_{c1}$ and $T_{c2}$ can be considered as collinear UUD phase. The NMR spectra within this phase has single-peaked shape. The phase boundaries for this phase marked on the phase diagram were obtained from the anomalies on the temperature dependencies of spin lattice relaxation time $T_{1}$. The temperature range of existence of this phase increases with field, consistent with theoretical expectations.\cite{Miyashita_1986}

\begin{figure}
\includegraphics[width=0.9\columnwidth,angle=0,clip]{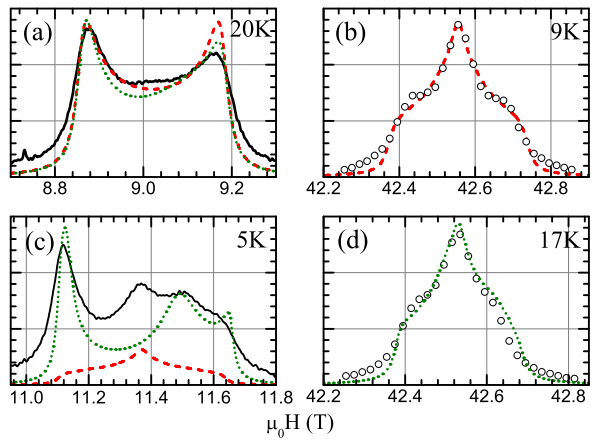}
\caption{(color online)
Fitting of NMR spectra (black lines and circles) within 3D and 2D incommensurate magnetic structures:
(a) - 3D, undistorted ($C=0$, red dashed line) and distorted ($C=0.013$, green dotted line), $M_1=M_2=2.1\mu_B$; (b) - 2D, undistorted ($C=0$), $M_1=M_2=2.4\mu_B$; (c) - 3D, distorted ($C=0.06$, green dotted line) and 2D, undistorted ($C=0$, red dashed line), $M_1=M_2=3.5\mu_B$; (d) - 2D, distorted ($C=0.015$), $M_1=M_2=2\mu_B$. Individual linewidth $\delta=18$~mT.
}
\label{fig:fitting}
\end{figure}

The fits to the experimental NMR spectra using suggested models are shown in Fig.\ref{fig:fitting}.
The NMR spectra measured in regions I and III can be fitted satisfactorily by 3D and 2D incommensurate magnetic structures. The observed asymmetry of the spectra can be described by the unharmonicity of helical structure discussed above (Fig.\ref{fig:fitting}a,d). The complicated shape of spectra observed within the hysteretic region II of the phase diagram can be fitted by superposition of the spectra from 2D and 3D incommensurate structures (Fig.\ref{fig:fitting}c).
The shape of NMR spectra is also dependent on the differences in spin-spin relaxation time $T_{2}$ along the spectrum lines. However, the measured $T_{2}$ along a spectrum for sampling spectra show that the $T_{2}$ correction of the spectra can change the relative intensity by not more than 15~\%.

Finally, we note that the local field on the nucleus measured by NMR experiment is formed by neighboring magnetic ions of few coordination spheres. It means that similar spectral shapes can also be obtained in the presence of corresponding short range correlations, static during the NMR experiment ($\gtrsim1$~ms).\cite{Comment}

\section{Conclusions}

The magnetic phase diagram of CuCrO$_2$ is studied with Cu NMR for $\vect{H}\parallel\vect{c}$. NMR experiments revealed the H-T regions where 2D/3D-planar spiral, UUD and paramagnetic phases are realized. The 2D-planar spiral phase realized in region III is very unusual. On one hand, according to NMR the usual 3D order is absent, but at the same time this phase possesses electric polarization.\cite{Zapf_2014} Symmetry analysis of magnetic properties of
CuCrO$_2$\cite{Marchenko_2014} allows one to explain these observations by a realization of 3D magnetic state with tensor order parameter. This phase can be classified as a polar nematic phase.

\acknowledgements

We thank V.~I.~Marchenko, N.~B\"{u}ttgen, V.~N.~Glazkov, A.~I.~Smirnov for stimulating
discussions. H.D.Z. thanks for the support from  NSF-DMR through award DMR-1350002. This work was supported by
Russian Foundation for Basic Research, Program of Russian Scientific Schools
(Grant 16-02-00688). Work at the National High Magnetic Field Laboratory is
supported by the NSF Cooperative Agreement No.~DMR-1157490, the State of
Florida, and the DOE.


\begin{thebibliography}{23}
\bibitem{Comment} This estimate we obtained from the fact that the effective field on each nucleus has to remain constant at least during experimentally measured time $T_1$. In the ordered phase $T_1\gtrsim1$ ms, see Fig.\ref{fig:t1}.
\end{thebibliography}
\end{document}